\documentclass[aps,pra,reprint,superscriptaddress]{revtex4-2}

\usepackage{mathrsfs}
\usepackage{graphicx}
\usepackage{epstopdf}
\usepackage{bm}
\usepackage{float}
\usepackage{fancyhdr}
\usepackage{longtable}
\usepackage{hyperref}
\usepackage{subfigure}
\usepackage{epsfig}
\usepackage{color,soul}
\usepackage{menukeys}
\usepackage{xcolor}
\usepackage{amssymb}
\usepackage{braket}
\usepackage{graphicx}
\usepackage{dcolumn}
\usepackage[mathscr]{eucal}
\usepackage{amsmath}
\usepackage{tabularx}
\usepackage{booktabs}
\usepackage{slashed}
\usepackage{mathtools,slashed}
\usepackage{axodraw2}
\usepackage[normalem]{ulem}

\makeatletter
\def\pslashed#1{\expandafter\ifx\csname psla@\string#1\endcsname\relax
	{\mathpalette{\sla@/00}{\phantom{#1}}}\else
	\csname psla@\string#1\endcsname\fi}
\def\declarepslashed#1#2#3#4#5{\expandafter\def\csname psla@\string#5\endcsname{#1{\mathpalette{\sla@{#2}{#3}{#4}}{\phantom{#5}}}}}
\makeatother
\declarepslashed{}{/}{.08}{0}{D}

\newcommand{\be}{\begin{equation}}
	\newcommand{\ee}{\end{equation}}
\newcommand{\bea}{\begin{eqnarray}}
	\newcommand{\eea}{\end{eqnarray}}
\newcommand{\ben}{\begin{enumerate}}
	\newcommand{\een}{\end{enumerate}}
\newcommand{\bde}{\begin{widetext}}
	\newcommand{\ede}{\end{widetext}}

\newcommand{\bc}{\begin{center}}
\newcommand{\ec}{\end{center}}

\begin{document}

\title{Majoron dark matter detection via hybrid magnon transmon qubit system}

\author{Le Bin Ho}
\email{ho.bin.le.e3@tohoku.ac.jp}
\affiliation{Department of Applied Physics, Graduate School of Engineering, Tohoku University, Sendai 980-8579, Japan}
\affiliation{Frontier Research Institute for Interdisciplinary Sciences, Tohoku University, Sendai 980-8578, Japan}

\author{Do Thi Huong}
\email{dthuong@iop.vast.vn}
\affiliation{Institute of Physics, Vietnam Academy of Science and Technology (VAST), No.10, Dao Tan, Giang Vo, Hanoi, Vietnam}
	
\date{\today}

\begin{abstract}
Ultralight Majorons are well-motivated dark-matter candidates that can couple to electron spins, generating an oscillating pseudo-magnetic field. We propose a hybrid magnon-qubit haloscope that exploits this interaction to search for Majoron dark matter. In our scheme, the Majoron field resonantly drives the Kittel mode of a ferrimagnetic yttrium iron garnet (YIG) sphere, producing a collective magnon response enhanced by the large spin ensemble. The resulting magnon population is transduced to a superconducting transmon qubit through a cavity-mediated dispersive interaction and detected using quantum-nondemolition Ramsey interferometry. Using experimentally demonstrated parameters for magnon-cavity-qubit systems, we derive the Majoron-induced signal, evaluate the projected sensitivity, and analyze the corresponding mass-scan strategy enabled by magnetic-field tuning, benchmarking the projected reach against existing laboratory and astrophysical constraints on the axion/Majoron-electron coupling. Our results establish hybrid magnon-qubit architectures as a promising quantum-sensing platform for searches for Majoron dark matter and, more generally, ultralight bosonic fields coupled to electronic spins.
\end{abstract}


\maketitle

\section{Introduction}
\label{sec:introduction}

Neutrino oscillation experiments have conclusively established that neutrinos are massive, providing compelling evidence for physics beyond the 
Standard Model (SM)~\cite{Fukuda1998,Ahmad2002}. Among the many extensions proposed to explain the origin of neutrino masses, scenarios
 involving Majorana neutrinos are particularly attractive because they naturally generate small neutrino masses through lepton-number-violating 
 mechanisms, most notably the seesaw framework~\cite{Minkowski1977,Mohapatra1980}. Such constructions often predict additional particles 
 beyond the SM spectrum, including light pseudo-Nambu--Goldstone bosons associated with the spontaneous breaking of global lepton number 
 symmetry~\cite{Chikashige1981,Gelmini1981,Huong2026}. The resulting boson, known as the Majoron, provides a natural link between neutrino mass generation and cosmology.

Although originally massless, the Majoron can acquire a small mass through explicit lepton-number-violating effects induced by ultraviolet physics, such as 
quantum gravity~\cite{Akhmedov1992,Rothstein1993}. In this case, it becomes a pseudo-Nambu--Goldstone boson that can remain stable on cosmological 
timescales and constitute all or a fraction of the observed dark matter abundance~\cite{Berezinsky1993,Gu2010,Frigerio2011}. Owing to its weak interactions and
 long lifetime, the Majoron has emerged as a well-motivated dark matter candidate in theories featuring spontaneously broken lepton number.

The possibility that the Majoron forms ultralight dark matter has motivated extensive efforts to identify observable signatures of its interactions with SM particles.
 Depending on the underlying realization, the Majoron can couple to neutrinos, photons, and charged fermions through derivative, Yukawa-like, or anomaly-induced 
 interactions~\cite{Srednicki:1985xd,Brune:2018sab, HerreroBrocal:2023}, leading to a rich phenomenology across laboratory experiments, astrophysical observations, and 
 cosmological probes~\cite{Lattanzi:2007ux}. Notably, an anomaly-induced coupling to photons makes the Majoron similar to the QCD axion. As thus, Majoron dark matter can be tested using existing axion dark matter searches and gravitational-wave interferometers operated as optical cavities~\cite{Liang2025,Obata2026}. In the ultralight regime, the Majoron dark matter background behaves as a coherently oscillating classical field
 whose spatial gradient acts effectively as a weak pseudo-magnetic field. Through its coupling to fermionic axial currents, this background can induce tiny oscillatory
  spin-dependent signals , which have been searched for using spin-precession and nuclear-magnetic-resonance-based techniques~\cite{Garcon2018,Graham2018Spin}, that are particularly suitable for precision quantum sensing experiments~\cite{Budker:2013hfa,Safronova:2017xyt,Carney:2019cio,Santoso2025}. Although ultralight dark matter may exist in nonclassical quantum states, these quantum effects are generally unobservable in realistic detectors. Therefore, a classical-field description provides an accurate framework for the analysis presented in this work~\cite{Bao2026}.

Among the various quantum sensing approaches, collective spin excitations in magnetic materials provide a particularly promising avenue for detecting spin-coupled 
dark matter. In ferrimagnetic systems, the collective dynamics of electron spins can be described in terms of magnons, with the uniform Kittel mode corresponding 
to the coherent precession of a macroscopic spin ensemble~\cite{Kittel:1948mn}. Because the dark matter field couples coherently to all participating spins, the resulting 
response benefits from a collective enhancement proportional to the size of the spin ensemble. This feature has stimulated considerable interest in magnon-based searches 
for ultralight bosonic dark matter, including axions, axion-like particles, and dark photons~\cite{Barbieri:2016ofm,Mitridate:2020vzg,Chigusa:2020gsk}.

Most existing magnon-based dark matter searches rely on classical microwave detection techniques, such as cavity transmission measurements and power 
spectroscopy~\cite{Flower:2018rms}. While these approaches have achieved impressive sensitivity, their performance is ultimately limited by thermal 
fluctuations, amplifier noise, and the standard quantum limit associated with linear microwave detection. 

Recent advances in superconducting quantum 
technologies have opened new opportunities for overcoming these limitations. In particular, hybrid magnon-qubit systems have emerged as a powerful interface
 between collective spin excitations and superconducting quantum circuits~\cite{Tabuchi2015,LachanceQuirion2020}. Through dispersive interactions, weak magnonic
  signals can be transduced into qubit observables and accessed using high-fidelity quantum non-demolition (QND) measurements, enabling operation in quantum-enhanced sensing regimes. 
This QND detection strategy has been experimentally demonstrated for axion-electron interactions. Using a ferrimagnetic sphere dispersively coupled to a superconducting qubit via a microwave cavity, Ikeda \textit{et al.} established the experimental viability of this detection scheme~\cite{Ikeda2022}.

In this work, we propose a hybrid magnon-qubit haloscope for the detection of ultralight Majoron dark matter as shown in Fig.~\ref{fig:setup}. The detector consists of a ferrimagnetic Kittel mode hosted
in an yttrium iron garnet (YIG) sphere and dispersively coupled to a superconducting transmon qubit through a microwave cavity. The oscillating Majoron field acts as an effective pseudo-magnetic drive
on the collective spin ensemble and coherently excites the Kittel mode when the resonance condition is satisfied. The resulting magnon population induces a measurable shift in the qubit
transition frequency, which is read out using Ramsey interferometry. We develop a theoretical framework describing the Majoron-induced magnon dynamics, the effective magnon-qubit interaction, and
the associated quantum measurement protocol. We further analyze the dominant noise sources and derive the resulting signal-to-noise ratio and projected sensitivity to the Majoron-electron axial coupling. We benchmark this projected sensitivity against existing laboratory constraints from the ferromagnetic haloscope QUAX-ae~\cite{Crescini2018,Crescini2020} and the magnon-QND search of Ikeda~\textit{et al.}~\cite{Ikeda2022}, as well as model-independent astrophysical and solar energy-loss bounds~\cite{Fleury2025,CapozziRaffelt2020,GondoloRaffelt2009}.
 Our results demonstrate that hybrid magnon-qubit architectures provide a promising quantum sensing platform for probing ultralight Majoron dark matter and offer a complementary approach to existing searches in the microwave-frequency regime.

The remainder of this paper is organized as follows. In Sec.~\ref{sec:magnon}, we investigate Majoron-induced magnon excitations
in a ferrimagnetic medium and derive the corresponding driven magnon response. In Sec.~\ref{sec:hybrid}, we develop the hybrid
magnon--qubit dynamics and obtain the effective dispersive description of the coupled system. Section~\ref{sec:measurement} presents
the QND detection strategy based on Ramsey interferometry and the associated signal generation mechanism.
In Sec.~\ref{sec:sensitivity}, we evaluate the projected sensitivity reach by analyzing the relevant noise sources and signal-to-noise ratio, and compare it with existing laboratory and astrophysical constraints on the axion/Majoron-electron coupling. Finally, our conclusions are summarized in Sec.~\ref{sec:conclusion}.

\section{Majoron-Induced Magnon Excitations}
\label{sec:magnon}
\subsection{Majoron dark matter as pseudo-magnetic field}
Throughout this work, we adopt natural units \( \hbar=c=k_B=1 \) unless stated otherwise. 
To establish the foundational signal generation mechanism, we first consider the scenario in which the Majoron constitutes all or a fraction of the Galactic dark matter abundance. 
Due to the extremely large occupation number of ultralight bosonic dark matter within a de Broglie volume, the Majoron field behaves as a classical coherent wave on laboratory scales. 
The corresponding field configuration can be expressed as \cite{Graham:2013gfa, Hui:2016ltb}
\begin{equation}
J(t,\mathbf{x}) = J_0\cos	\!\left(m_J t	- \mathbf{k}\cdot\mathbf{x}	+\theta_J\right),	\label{eq:Jfield}
\end{equation}
where \(m_J\) denotes the Majoron mass, \(\mathbf{k}\) is the dark matter momentum, and \(\theta_J\) is an arbitrary initial phase. 
For a virialized Galactic halo, the momentum satisfies \(|\mathbf{k}|=m_Jv_{\rm DM}\), with a characteristic virial velocity \(v_{\rm DM}\sim10^{-3}\).
The field amplitude is fixed by the local dark matter energy density through \(\rho_{\rm DM} = \frac12 m_J^2 J_0^2\), which implies
\begin{equation}
m_JJ_0 = \sqrt{2\rho_{\rm DM}}. \label{eq:JAmplitude} \end{equation}
Throughout this work, we adopt the commonly used estimate of the local dark matter density,
\(\rho_{\rm DM}\simeq0.4~{\rm GeV/cm^3}\), consistent with astrophysical determinations \cite{Read:2014qva}. Consequently, the spatial gradient of the Majoron field is given by
\begin{equation}
\nabla J =\mathbf{k}\,J_0\sin\!\left(m_J t-\mathbf{k}\cdot\mathbf{x}+\theta_J	\right),
\end{equation}
with a characteristic macroscopic amplitude \(|\nabla J| \sim m_Jv_{\rm DM}J_0 = v_{\rm DM} \sqrt{2\rho_{\rm DM}}\).
The coherence time of the Majoron background is determined by the dark
matter velocity dispersion,
\begin{equation}
\tau_J \simeq \frac{1}{m_Jv_{\rm DM}^2},
\end{equation}
which exceeds the intrinsic oscillation period by a factor
\(v_{\rm DM}^{-2}\sim10^6\)
\cite{Graham:2013gfa,Hui:2016ltb}.  As a result, the Majoron field behaves as a highly coherent classical source during each experimental interrogation cycle.

Rather than committing to a specific ultraviolet completion,
we adopt a model-independent effective-field-theory description
of the Majoron interaction with electrons \cite{HerreroBrocal:2023},
\begin{equation}
\mathcal L_{Jee} = \dfrac{c_e^A}{f_J}(\partial_\mu J) \bar e \gamma^\mu\gamma^5 e,
\end{equation}
where \(f_J\) denotes the Majoron symmetry-breaking scale and \(c_e^A\) parameterizes the effective axial coupling strength.
In the non-relativistic limit, the dominant contribution arises from the coupling between the Majoron spatial gradient and the electron spin density ~\cite{HerreroBrocal:2023},
\begin{equation}	\mathcal{L}_{Jee} \simeq \dfrac{2c_e^A}{f_J} (\nabla J)\cdot\mathbf{S}_e.	\label{eq:SpinCoupling}
\end{equation}
For a polarized spin ensemble occupying a volume \(V\), the corresponding effective interaction Hamiltonian reduces to
\begin{equation}
H_A = - \dfrac{2c_e^A}{f_J} (\nabla J) \cdot \mathbf{S}_{\rm tot}, \label{eq:HA}
\end{equation}
where \(\mathbf{S}_{\rm tot} = \int_V \mathrm{d}^3x\,\mathbf{S}_e(\mathbf{x})\) denotes the collective macro-spin operator of the medium.
Equation~(\ref{eq:HA}) can be directly mapped onto the interaction of the spin ensemble with an effective oscillating pseudo-magnetic field \cite{Graham:2013gfa,Budker:2013hfa}. 
Comparing with the conventional electronic Zeeman Hamiltonian
\(H_Z=-g_e\mu_B\,\mathbf B\cdot\mathbf S_{\rm tot}\),
one identifies the Majoron-induced effective pseudo-magnetic field as
\begin{equation}
\mathbf B_{\rm eff}(t)= \dfrac{2c_e^A}{g_e\mu_Bf_J} \nabla J(t). \label{eq:Beff}
\end{equation}
Using Eq.~(\ref{eq:Jfield}), the effective field oscillates coherently at the Majoron Compton frequency,
\begin{equation}
\mathbf B_{\rm eff}(t)=\mathbf B_{\rm eff}^{(0)}\sin\!\left(m_J t-\mathbf k\!\cdot\!\mathbf x+\theta_J \right),
\end{equation}
with amplitude
\begin{equation}
B_{\rm eff}^{(0)} = \dfrac{2c_e^A} {g_e\mu_Bf_J} v_{\rm DM}\sqrt{2\rho_{\rm DM}}. \label{eq:BeffAmplitude}
\end{equation}
The ultralight Majoron background therefore acts as a weak oscillating pseudo-magnetic field that couples coherently to polarized electron spins.

\subsection{Coupling to the Kittel magnon mode in YIG}

To convert the extremely weak pseudo-magnetic field generated by dark matter into a measurable quantum signal, we exploit collective spin excitations in a macroscopic magnetic medium. 
Specifically, we consider a spherical YIG sample placed in a static uniform magnetic field \(\mathbf{B}_0 = B_0 \hat{\mathbf z}\).
The equilibrium ground state can be described, at the level of the
uniform Kittel mode, by an effective saturated macro-spin
\(S_{\rm tot}=N_sS\), where \(N_s\) denotes the effective number of
participating Fe\(^{3+}\) spin sites and \(S=5/2\) is the spin quantum number associated with the \(\text{Fe}^{3+}\) ions in the YIG lattice.
Low-energy collective excitations above this ferromagnetic ground state are described in terms of quasiparticle excitations known as magnons~\cite{Cherepanov1993}. 
For the uniform ferromagnetic resonance (FMR) mode, commonly referred to as the
Kittel mode, the collective spin operators can be mapped onto bosonic
creation and annihilation operators through the Holstein--Primakoff
transformation~\cite{Holstein1940}:
\begin{equation}
	\hat S_{\rm tot}^{+} \simeq \sqrt{2N_sS}\,\hat a, \
	\hat S_{\rm tot}^{-} \simeq \sqrt{2N_sS}\,\hat a^\dagger, \
	\hat S_{\rm tot}^{z} = N_sS - \hat a^\dagger\hat a.
\end{equation}
This linearized bosonic approximation is strictly valid in the low-excitation regime \(\langle\hat a^\dagger\hat a\rangle \ll N_sS\), which is rigorously satisfied for the faint dark matter drives considered in this work.
The unperturbed free Hamiltonian of the Kittel mode then reduces to
\begin{equation}
H_m = \omega_K \hat a^\dagger \hat a. \label{eq:Hm}
\end{equation}

Neglecting crystalline anisotropy fields, the isotropic demagnetization
tensor of a spherical sample, \(N_x=N_y=N_z=1/3\), reduces the Kittel
frequency to the Zeeman form
\(\omega_K\simeq\gamma_e B_0\),
where \(\gamma_e=g_e\mu_B\) in natural units \cite{Kittel:1948mn}.

Having established the bosonic description of the magnetic medium, we now evaluate the coupling between the oscillating Majoron background and the Kittel magnon mode. 
For maximal transverse driving, we take the Majoron-field gradient to be oriented perpendicular to the equilibrium magnetization axis, \(\nabla J(t) = (\nabla J)_x \hat{\mathbf{x}}\). 
The effective interaction Hamiltonian then becomes
\begin{equation}
H_A = -\dfrac{c_e^A}{f_J} (\nabla J)_x \left( \hat S_{\rm tot}^{+}+ \hat S_{\rm tot}^{-} \right).
\end{equation}
Applying the linearized Holstein--Primakoff transformation yields a time-dependent bosonic drive acting on the Kittel mode,
\begin{equation}
H_A = \Omega_J(t) \left( \hat a + \hat a^\dagger \right), \label{eq:MajoronDrive}
\end{equation}
where the time-dependent drive profile is given by
\begin{equation}
\Omega_J(t) = \Omega_J^{(0)} \sin \left( m_J t+\theta_J \right),
\end{equation}
with the collective drive amplitude defined as
\begin{equation}
\Omega_J^{(0)} = \dfrac{c_e^A}{f_J} v_{\rm DM}\sqrt{2\rho_{\rm DM}} \sqrt{2N_sS}. \label{eq:OmegaJ}
\end{equation}
Equation~(\ref{eq:OmegaJ}) explicitly exhibits the characteristic collective enhancement 
factor \(\sqrt{N_s}\), arising from the coherent participation of all spins in the uniform Kittel mode. 
Consequently, a weak Majoron-induced perturbation acting on individual electron spins is collectively amplified 
by the macroscopic spin ensemble, in direct analogy with the collective spin enhancement widely exploited
in cavity-magnonic systems~\cite{Ikeda2022,Tabuchi2015,Ren2025,LachanceQuirion2020}.

\section{Hybrid Magnon--Qubit Dynamics}
\label{sec:hybrid}
\begin{figure}[t]
\centering
\includegraphics[width=\linewidth]{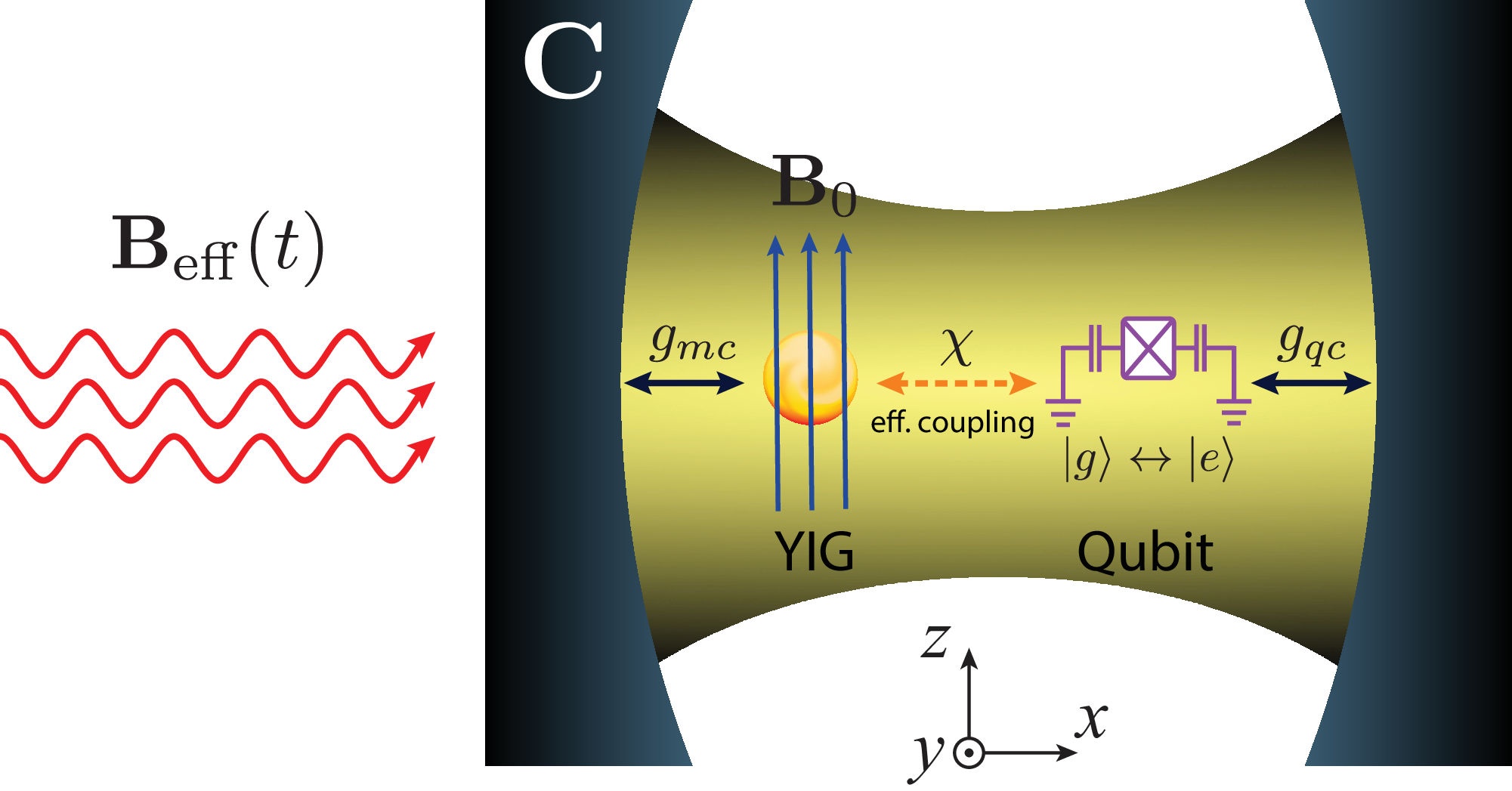}
\caption{Sketch of the proposed hybrid magnon--qubit haloscope for Majoron dark matter detection. A spherical YIG sample, biased by a static field $\mathbf B_0$, supports a Kittel magnon mode ($\hat a,\hat a^\dagger$) driven by the effective pseudo-magnetic field $\mathbf{B}_{\rm eff}(t)$ induced by the oscillating Majoron dark matter field $J(t,\mathbf{x})$. The YIG sphere and a superconducting transmon qubit are simultaneously coupled to a 3D superconducting microwave cavity, with coupling strengths $g_{mc}$ and $g_{qc}$, respectively, giving rise to an effective dispersive magnon--qubit interaction $\chi$. The entire setup is operated inside a dilution refrigerator at $T\sim20$~mK to ensure near-quantum-limited operation. The Majoron-induced magnon population is imprinted onto the qubit phase and extracted through a QND Ramsey readout, yielding the measured signal $\langle\hat\sigma_z\rangle$.}
\label{fig:setup}
\end{figure}

To enable quantum-limited detection of the Majoron-induced magnon response, we consider a hybrid quantum platform consisting of a spherical 
YIG resonator and a superconducting transmon qubit coupled via a three-dimensional microwave cavity as shown in Fig.~\ref{fig:setup}. The cavity serves as a quantum bus mediating
coherent interactions between the two subsystems, while the transmon qubit provides a high-fidelity QND probe of the magnon population.
This cavity-mediated architecture has been extensively investigated in both experimental and theoretical studies~\cite{Huebl2013, Tabuchi2015,Ren2025,LachanceQuirion2020,Ikeda2022}.

\subsection{Effective dispersive Hamiltonian}
The total system Hamiltonian is given by
\begin{align}
H_{\rm sys} &= \omega_K \hat a^\dagger \hat a +\omega_c \hat c^\dagger \hat c +\frac{\omega_q}{2}\hat\sigma_z
+ g_{mc}\left(\hat a^\dagger \hat c+\hat a\hat c^\dagger\right) \notag \\
&+ g_{qc}\left(\hat\sigma_+\hat c+\hat\sigma_-\hat c^\dagger\right) +\Omega_J(t)\left(\hat a+\hat a^\dagger\right),
\label{eq:Hsys}
\end{align}
where $\hat a$ and $\hat c$ are the annihilation operators of the Kittel magnon and cavity photon modes, respectively. 
The transmon qubit is described by the Pauli operator $\hat\sigma_z$ and the ladder operators $\hat\sigma_\pm$.
 The parameters $g_{mc}$ and $g_{qc}$ denote the vacuum magnon-cavity and qubit-cavity coupling strengths, respectively.

We operate in the dispersive regime where direct energy exchange is strongly suppressed, $g_{mc(qc)} \ll |\Delta_{mc(qc)}|$, 
with the subsystem detunings defined as $\Delta_{mc}=\omega_K-\omega_c$ and $\Delta_{qc}=\omega_q-\omega_c$. 
To derive the effective low-energy description, we apply a unitary Schrieffer-Wolff
 transformation, $H_{\rm disp}^{(1)} = e^{\hat V} H_{\rm sys} e^{-\hat V}$, generated by the anti-Hermitian operator
\begin{equation}
\hat V = \frac{g_{qc}}{\Delta_{qc}} \left( \hat\sigma_+\hat c - \hat\sigma_-\hat c^\dagger \right) 
+ \frac{g_{mc}}{\Delta_{mc}} \left( \hat a^\dagger\hat c - \hat a \hat c^\dagger \right).
\end{equation}
Expanding perturbatively to second order and adiabatically eliminating the 
cavity vacuum subspace yields the effective magnon-qubit Hamiltonian:

\begin{align}
H_{\rm disp}^{(1)}  \simeq{}\tilde{\omega}_K^{(1)} \hat a^\dagger\hat a
	&+\tilde{\omega}_q^{(1)} 
	\hat\sigma_z + J_{mq}\left( \hat a\hat\sigma_+ +\hat a^\dagger\hat\sigma_- \right)\notag \\
	&+\Omega_J(t)
	\left(\hat a+\hat a^\dagger
\right).
\end{align}
where the cavity-mediated magnon-qubit exchange coupling is
\begin{equation}
J_{mq} = \frac{g_{mc}g_{qc}}{2} 	\left(
	\frac{1}{\Delta_{mc}} + \frac{1}{\Delta_{qc}}\right), \label{eq:Jmq}
\end{equation}
and we have defined the dispersively renormalized frequencies
\begin{equation}
\tilde{\omega}_K^{(1)}  =
\omega_K+\frac{g_{mc}^{2}}{\Delta_{mc}},	\qquad 	\tilde{\omega}_q^{(1)} 
	=	\omega_q+\frac{g_{qc}^{2}}{\Delta_{qc}}.
\end{equation}
Equation~\eqref{eq:Jmq} describes an effective coupling between the 
Kittel magnon mode and the transmon qubit mediated by virtual cavity photons.

To obtain a QND readout Hamiltonian, we further assume that the 
effective magnon-qubit system operates in the dispersive regime,
\begin{equation}
	|J_{mq}|\ll |\Delta_{mq}|,
	\qquad
	\Delta_{mq}
	=
	\tilde{\omega}_q-\tilde{\omega}_K.
\end{equation}
Under this condition, the effective exchange interaction can be perturbatively 
eliminated through a second Schrieffer-Wolff transformation,
\begin{equation}	H_{\rm eff} = e^{\hat S} H_{\rm disp}^{(1)}e^{-\hat S},\qquad\hat S
	=\frac{J_{mq}}{\Delta_{mq}}\left(\hat a\hat\sigma_+-\hat a^\dagger\hat\sigma_-\right).
\end{equation}
Expanding the transformed Hamiltonian to second order in
$J_{mq}/\Delta_{mq}$ generates the dispersive cross-Kerr interaction,
\begin{equation}\chi=\frac{J_{mq}^{2}}{\Delta_{mq}}
=\frac{g_{mc}^{2}g_{qc}^{2}}{4\Delta_{mq}}\left(\frac{1}{\Delta_{mc}}+\frac{1}{\Delta_{qc}}\right)^2,\label{eq:chi}
\end{equation}
while the accompanying qubit Lamb shift,
$\chi\hat\sigma_z/2$,
is absorbed into the renormalized qubit frequency.  The resulting low-energy Hamiltonian is therefore
\begin{equation}
H_{\rm eff}\simeq \tilde{\omega}_K \hat a^\dagger \hat a+\frac{\tilde{\omega}_q}{2}\hat\sigma_z+\chi \hat a^\dagger \hat a\,\hat\sigma_z
+\Omega_J(t)\left(\hat a+\hat a^\dagger	\right),\label{eq:Heff}
\end{equation}
where the renormalized frequencies $\tilde{\omega}_K=\tilde{\omega}_K^{(1)}$ and $\tilde{\omega}_q=\tilde{\omega}_q^{(1)}+ \chi$ 
include all dispersive frequency shifts generated by the two successive Schrieffer-Wolff transformations.
The cross-Kerr interaction,
\(H_{\rm Kerr}=\chi \hat a^\dagger \hat a\,\hat\sigma_z,\)
shifts the qubit transition frequency conditionally on the magnon occupation number. For a magnon population
\(\langle\hat n\rangle=\langle\hat a^\dagger\hat a\rangle,\) the effective qubit transition frequency is
\begin{equation}\omega_q^{\rm eff} =\tilde{\omega}_q+2\chi\langle\hat n\rangle,	\label{eq:QubitShift}
\end{equation}
which forms the basis of the Ramsey QND readout developed in the next section.

\begin{figure*}[t]
\centering
\includegraphics[width=\textwidth]{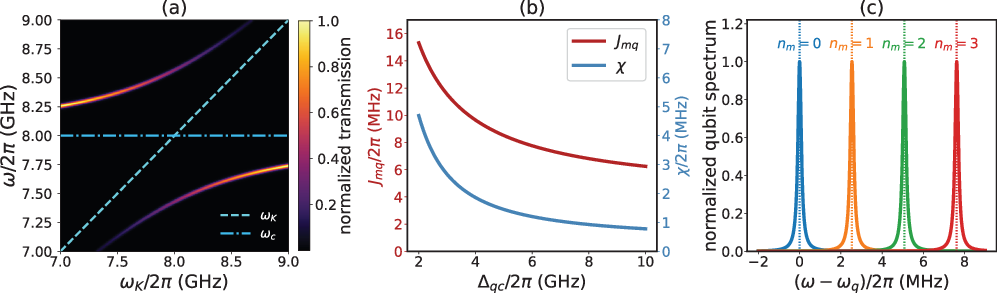}
\caption{Characterization of the hybrid magnon-qubit system. (a) Normalized cavity transmission $|S_{21}(\omega)|^2$ as a 
	function of probe frequency $\omega/2\pi$ and Kittel-mode frequency $\omega_K/2\pi$, showing the avoided crossing of the two hybridized normal-mode branches expected in the strong magnon-cavity coupling regime. (b) Effective magnon--qubit coupling $J_{mq}/2\pi$ (left axis) and dispersive cross-Kerr interaction $\chi/2\pi$ (right axis) versus qubit--cavity detuning $\Delta_{qc}/2\pi$. Both interactions decrease with increasing detuning, illustrating the trade-off between interaction strength and dispersive operation. (c) Qubit spectra for fixed magnon occupations $n_m=0,1,2,3$, with resonances shifted by $2\chi n_m$. The large separation between adjacent peaks enables single-magnon-resolved readout.}
\label{fig:hybrid}
\end{figure*} 

\subsection{Hybrid-system characterization}
Figure~\ref{fig:hybrid} summarizes the key characteristics of the proposed hybrid magnon--qubit system. Panel~(a) shows the cavity transmission spectrum. Following the standard input--output formalism~\cite{Gardiner1985}, the transmission amplitude is derived as,
\begin{equation}
S_{21}(\omega)=
\left[
i(\omega_c-\omega)+\frac{\kappa_c}{2}
+\frac{g_{mc}^2}{i(\omega_K-\omega)+\kappa_m/2}
\right]^{-1},
\label{eq:S21}
\end{equation}
where $\omega$ is the frequency of the microwave probe tone used to interrogate the cavity, $\kappa_c$ is the cavity's total photon linewidth (energy decay rate), and $\kappa_m$ is the Kittel-mode (magnon) linewidth, i.e.\ the intrinsic magnon damping rate. Following the Gilbert damping model established for high-quality YIG spheres~\cite{Tabuchi2015,LachanceQuirion2020}, the magnon linewidth grows linearly with the Kittel-mode frequency,
\begin{equation}
\kappa_m(f)=2\pi f\,\alpha_G,
\label{eq:GilbertDamping}
\end{equation}
with $\alpha_G=1.9\times10^{-4}$. Near the bare cavity resonance, this gives $\kappa_m/2\pi\approx1.5~\mathrm{MHz}$ at $f=8~\mathrm{GHz}$. We evaluate at $\omega_c/2\pi=8~\mathrm{GHz}$ and $\kappa_c/2\pi=2.5~\mathrm{MHz}$, representative of the TE$_{10p}$ cavity modes reported in Refs.~\cite{Tabuchi2015,LachanceQuirion2020}, and $g_{mc}=g_0\sqrt{N_s}$ with $g_0/2\pi\approx17.9~\mathrm{mHz}$~\cite{Tabuchi2015}, giving $g_{mc}/2\pi\approx567~\mathrm{MHz}$ at $N_s=1\times10^{21}$.  As the Kittel-mode frequency $\omega_K$ is tuned through the cavity resonance $\omega_c$, the clear avoided crossing with a splitting of approximately $2g_{mc}$ confirms that the magnon and cavity modes operate in the strong-coupling regime ($g_{mc}\gg \kappa_c,\kappa_m$), consistent with previous experiments~\cite{Tabuchi2015,LachanceQuirion2020,Ren2025}.

Figure~\ref{fig:hybrid}(b) shows the effective magnon-qubit coupling $J_{mq}$ and dispersive cross-Kerr interaction $\chi$ 
as functions of the qubit cavity detuning $\Delta_{qc}$, at the fixed operating detuning $\Delta_{mc}/2\pi=5.7~\mathrm{GHz}$, chosen 
to satisfy $g_{mc}\ll\Delta_{mc}$, with $g_{qc}/2\pi=80~\mathrm{MHz}$, a typical transmon 3D cavity coupling strength in circuit 
QED~\cite{Tabuchi2015,LachanceQuirion2020}, and $\Delta_{mq}/2\pi=50~\mathrm{MHz}$. Both quantities decrease as the qubit is 
moved farther from the cavity resonance, illustrating the trade-off between interaction strength and maintaining a well-defined 
dispersive regime. At the operating point $\Delta_{qc}=\Delta_{mc}$, Eq.~\eqref{eq:chi} gives the dispersive shift as $\chi/2\pi\approx1.27~\mathrm{MHz}.$

Figure~\ref{fig:hybrid}(c) shows the qubit spectrum for magnon occupations $n_m=0,1,2,3$ at  $\chi/2\pi = 1.27~\mathrm{MHz}$. Each spectrum is described by a Lorentzian lineshape,
\begin{equation}
L_n(\omega)=
\frac{(\kappa_q/2)^2}
     {(\omega-\omega_q-2\chi n_m)^2+(\kappa_q/2)^2},
\label{eq:QubitLineshape}
\end{equation}
where the qubit resonance is shifted by $2\chi$ for each additional magnon, and $\kappa_q/2\pi\approx0.20~\mathrm{MHz}$ is 
the qubit linewidth~\cite{LachanceQuirion2020}. The peak separation is much larger than the qubit linewidth ($2\chi \gg \kappa_q$), allowing 
neighboring magnon-number states to be spectroscopically resolved.
\section{Quantum Measurement Protocol}
\label{sec:measurement}
\subsection{Steady-state magnon response}
To describe the driven magnon dynamics under realistic experimental conditions, we treat the system within the framework of 
an open quantum system. We first transform Eq.~\eqref{eq:Heff} into a frame rotating simultaneously with 
the Majoron oscillation frequency and the qubit transition frequency through the unitary transformation
\begin{equation}
	U_{mq}(t)
	=
	\exp\!\left[
	im_Jt\,\hat a^\dagger\hat a
	+
	i\frac{\tilde{\omega}_q}{2}t\,\hat\sigma_z
	\right].
\end{equation}
The transformed Hamiltonian is
\begin{equation}
	H_{mq}^{(R)}
	=
	U_{mq}H_{\rm eff}U_{mq}^\dagger
	-
	iU_{mq}
	\frac{\partial U_{mq}^\dagger}{\partial t}.
\end{equation}
Using
\(
U_{mq}\hat aU_{mq}^\dagger=e^{-im_Jt}\hat a
\)
and
\(
U_{mq}\hat a^\dagger U_{mq}^\dagger=e^{im_Jt}\hat a^\dagger,
\)
the free qubit evolution is exactly removed, while the Majoron drive contains both time-independent and rapidly oscillating components proportional to $e^{\pm2im_Jt}$. Within the rotating-wave approximation, the rapidly oscillating terms are neglected, yielding
\begin{equation}
	H_{mq}^{(R)}
	\simeq
	\delta\tilde{\omega}_K
	\hat a^\dagger\hat a
	+
	\chi \hat a^\dagger\hat a\,\hat\sigma_z
	+
	\frac{\Omega_J^{(0)}}{2}
	\left(
	\hat a+\hat a^\dagger
	\right),
	\label{eq:HR}
\end{equation}
where
\begin{equation}
	\delta\tilde{\omega}_K
	=
	\tilde{\omega}_K-m_J
\end{equation}
is the detuning between the renormalized Kittel-mode frequency and the Majoron oscillation frequency.
The cross-Kerr interaction is retained in Eq.~(\ref{eq:HR}) since it provides the dispersive coupling required for the qubit readout. In deriving the coherent magnon response, we treat the qubit as a weak, non-invasive probe and neglect its backaction on the magnon dynamics. Including intrinsic magnon dissipation at a rate \(\kappa_m\), which is the linewidth of the Kittel mode, the coherent magnon amplitude \(\alpha\equiv\langle\hat a\rangle\) obeys the semi-classical equation of motion,
\begin{equation}
	\frac{\mathrm d\alpha}{\mathrm dt}
	=
	-\left(
	i\delta\tilde{\omega}_K
	+
	\frac{\kappa_m}{2}
	\right)\alpha
	-i\frac{\Omega_J^{(0)}}{2}.
	\label{eq:AlphaEquation}
\end{equation}
The competition between the continuous Majoron drive and intrinsic magnon dissipation causes the transient dynamics to decay on a timescale of order \(2/\kappa_m\). Consequently, after a sufficiently long evolution, the system approaches a steady state in the rotating frame, where the coherent magnon amplitude becomes time independent,
\begin{equation}
	\frac{\mathrm d\alpha}{\mathrm dt}=0.
\end{equation}
The steady-state solution of Eq.~(\ref{eq:AlphaEquation}) is therefore
\begin{equation}
\alpha_{\rm ss} = -\frac{i\Omega_J^{(0)}/2} {\kappa_m/2+i\delta\tilde{\omega}_K},\label{eq:AlphaSS}
\end{equation}
which yields the coherent steady-state magnon occupation number:
\begin{equation}
	\langle \hat n \rangle = |\alpha_{\rm ss}|^2 = \frac{\left(\Omega_J^{(0)}/2\right)^2}{\delta\tilde{\omega}_K^2 + \left(\kappa_m/2\right)^2}.
	\label{eq:MagnonPopulation}
\end{equation}
Equation~\eqref{eq:MagnonPopulation} exhibits the characteristic Lorentzian response profile of a driven damped harmonic oscillator. 
The signal conversion efficiency is maximized at exact resonance, \(\delta\tilde{\omega}_K = 0\) (\(m_J \simeq \tilde{\omega}_K\)), where 
the oscillating Majoron background is most efficiently converted into a coherent Kittel-mode magnon population. 
This resonantly enhanced magnon population constitutes the primary dark-matter-induced signal in our proposal.

\subsection{Ramsey Interferometry}

The steady-state magnon population obtained above constitutes the primary physical signal generated by the Majoron field.
To detect this signal experimentally, it is transduced into a measurable superconducting-qubit phase through 
the dispersive cross-Kerr interaction.  Since the coherent magnon population has already reached its steady state before 
the Ramsey interrogation begins, the magnon occupation remains effectively constant throughout the free-evolution interval. 
Consequently, the dispersive interaction acts as a static qubit frequency shift during the Ramsey sequence.

The Ramsey protocol consists of two resonant $\pi/2$ microwave pulses separated by a free-evolution interval $t_R$, 
referred to as the Ramsey interrogation time. The first resonant $\pi/2$ pulse prepares the qubit in an equal coherent superposition of its ground and excited states,
\begin{equation}
|\psi(0)\rangle = \frac{|g\rangle+|e\rangle}{\sqrt{2}},
\end{equation}
up to an overall phase convention. During the subsequent free evolution, the qubit evolves under the dispersive interaction with the steady-state magnon 
occupation. Since the free qubit precession has already been removed by the rotating-frame transformation, the effective Hamiltonian governing the Ramsey evolution reduces to
\begin{equation}
H_q^{(R)} = \chi \langle\hat{n}\rangle \hat{\sigma}_z.	\label{eq:HqRamsey}
\end{equation}
The corresponding unitary evolution during the interrogation interval is
\begin{equation}
U_R(t_R) = \exp\left(-iH_q^{(R)}t_R\right) = \exp\left(-i\chi\langle\hat{n}\rangle t_R\hat{\sigma}_z\right),
	\label{eq:RamseyUnitary}
\end{equation}
so that the ground and excited states, being eigenstates of $\hat{\sigma}_z$ with eigenvalues $\mp1$,
acquire opposite dynamical phases. Consequently, the accumulated relative phase is
\begin{equation}
\phi_{\rm sig} = 2\chi\langle\hat{n}\rangle t_R,\label{eq:RamseyPhase}
\end{equation}
which is directly proportional to the Majoron-induced steady-state magnon occupation.

Finally, a second resonant $\pi/2$ pulse rotates the qubit back to the readout basis. Since the dispersive interaction modifies only the 
relative phase between the $|g\rangle$ and $|e\rangle$ components without changing their populations, the second pulse recombines the 
two coherent components of the qubit state and converts the accumulated phase into a measurable population difference. 
The phase of this pulse serves as an experimentally tunable reference phase, allowing the Ramsey fringe to be measured at an arbitrary operating point. Consequently, the excited-state probability is
\begin{equation}
	P_e = \frac{1}{2} \left[ 1 + \cos\left( \phi_{\rm sig} + \varphi_0 \right) \right],
	\label{eq:RamseyProbability}
\end{equation}
where $\varphi_0$ is the phase of the second $\pi/2$ pulse.

For optimal sensitivity to weak signals, the Ramsey sequence is operated at the quadrature point by choosing $\varphi_0 = -\pi/2$, for 
which the Ramsey signal is maximally sensitive to small phase variations. The measured signal, expressed as the expectation value of the Pauli operator,
\begin{equation}
	\langle\hat{\sigma}_z\rangle = 2P_e - 1 = \sin(\phi_{\rm sig}),
\end{equation}
reduces, in the weak-signal regime ($\phi_{\rm sig} \ll 1$), to
\begin{equation}
	\langle\hat{\sigma}_z\rangle \simeq \phi_{\rm sig} = 2\chi\langle\hat{n}\rangle t_R =\frac{2\chi \left(\Omega_J^{(0)}/2\right)^2}{\delta\tilde{\omega}_K^2 + \left(\kappa_m/2\right)^2} t_R,
	\label{eq:RamseyLinear}
\end{equation}
which provides a linear transduction of the Majoron-induced magnon occupation into the experimentally measured qubit signal through the standard dispersive readout.

The Ramsey sequence therefore realizes a QND measurement of the Majoron-induced magnon occupation. During the free-evolution interval, the dispersive interaction imprints the magnon occupation onto the relative phase of the qubit state, while the final $\pi/2$ pulse maps this phase information onto the qubit population without exchanging excitations with the magnon mode, thereby enabling high-sensitivity readout while preserving the magnon occupation throughout the measurement process.

\begin{figure*}[t]
\centering
\includegraphics[width=0.8\textwidth]{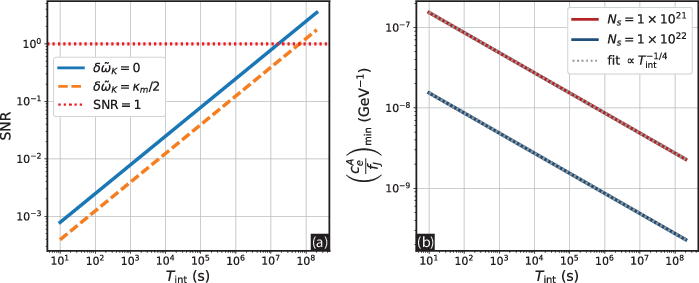}
\caption{(a) Signal-to-noise ratio versus integration time $T_{\rm int}$, at $\delta\tilde\omega_K=0$ and at $\delta\tilde\omega_K=\kappa_m/2$, for a reference coupling $c_e^A/f_J=4.3\times10^{-9}~\mathrm{GeV}^{-1}$. The dotted line marks ${\rm SNR}=1$. (b) Minimum detectable coupling $(c_e^A/f_J)_{\rm min}$ versus $T_{\rm int}$ at $\delta\tilde\omega_K=0$, for $N_s=1\times10^{21}$ and $N_s=1\times10^{22}$, together with the $T_{\rm int}^{-1/4}$ power-law guide (dotted) confirming the scaling as shown in the main text.}
\label{fig:snr}
\end{figure*}

\section{Sensitivity Projections}
\label{sec:sensitivity}
\subsection{Signal-to-Noise ratio}
Having established the Ramsey response in the linear regime, we now estimate the projected sensitivity of the proposed detector
by comparing the Majoron-induced qubit signal with the dominant experimental noise sources. For weak Majoron signals, the measured Ramsey response is
\begin{equation}
S \equiv \langle\hat{\sigma}_z\rangle \simeq 2\chi \langle\hat n\rangle t_R. \label{eq:Signal}
\end{equation}
The Ramsey experiment is repeated \(N_{\rm meas}\) times during a total integration time
\begin{equation} T_{\rm int} = N_{\rm meas} t_{\rm cycle}, \end{equation}
where \(t_{\rm cycle}\) is the duration of a single Ramsey sequence.  Assuming statistically independent measurements, the quantum projection noise (shot noise) is
\begin{equation} \sigma_{\rm shot} = \frac1{\sqrt{N_{\rm meas}}} = \sqrt{\frac{t_{\rm cycle}} {T_{\rm int}}}.	\label{eq:ShotNoise}
\end{equation}
Additional noise contributions include thermal magnon fluctuations and measurement back-action,
\begin{equation}
\sigma_{\rm tot}^2 = \sigma_{\rm shot}^2+\sigma_{\rm thermal}^2+\sigma_{\rm back}^2.\label{eq:Noise}
\end{equation}
For dilution-refrigerator temperatures(\(T\simeq20~{\rm mK}\)) and GHz-frequency magnons, the thermal occupation satisfies \(n_{\rm th}\ll1,\)
while the dispersive Ramsey protocol realizes a QND readout that strongly suppresses measurement back-action.
Consequently,
\begin{equation} \sigma_{\rm tot} \simeq \sigma_{\rm shot}.
\end{equation}
The signal-to-noise ratio is therefore
\begin{equation}
{\rm SNR} =\frac{S} {\sigma_{\rm shot}}=2\chi \langle\hat n\rangle t_R \sqrt{\frac{T_{\rm int}} {t_{\rm cycle}}}. \label{eq:SNR}
\end{equation}
Equation~(\ref{eq:Signal}) shows that the Ramsey signal increases linearly with the interrogation time \(t_R\). 
A longer interrogation time therefore improves the projected detector sensitivity by allowing a larger phase accumulation.
In practice, however, \(t_R\) is limited by two independent considerations: it must remain shorter than the qubit coherence time \(T_2\) 
to preserve the Ramsey fringe visibility, and the accumulated phase should satisfy \(\phi_{\rm sig}\lesssim1\) to ensure operation in the linear-response regime. Accordingly, we choose
\[t_R=\min\left(T_2,\frac{1}{2\chi\langle\hat n\rangle}\right),\]
which maximizes the measurable signal while maintaining both qubit coherence and linear detector response.
\subsection{Minimum detectable Majoron coupling}
The detector sensitivity is conventionally defined by the minimum Majoron-induced drive amplitude that yields a unit signal-to-noise ratio.
Setting \({\rm SNR}=1, \) Eq.~(\ref{eq:SNR}) gives
\begin{equation}
\Omega_{J,\min}^{(0)} = \kappa_m \left( \frac1{2\chi t_R}\sqrt{\frac{t_{\rm cycle}}{T_{\rm int}}} \right)^{1/2}.\label{eq:OmegaMin}
\end{equation}
It leads to the constraint
\begin{equation}
\left(\dfrac{c_e^A}{f_J} \right)_{\rm min}
=\frac{\kappa_m} {v_{\rm DM} \sqrt{2\rho_{\rm DM}} \sqrt{2N_sS}}\left(
\frac1{2\chi t_R}\sqrt{\frac{t_{\rm cycle}}{T_{\rm int}}}\right)^{1/2}.
\label{eq:Sensitivity}
\end{equation}
Equation~(\ref{eq:Sensitivity}) explicitly illustrates the scaling of the projected detector sensitivity.
The reachable coupling improves with stronger dispersive interaction ($\chi$), longer Ramsey
 interrogation times ($t_R$), and larger integration times ($T_{\rm int}$).
Furthermore, the collective enhancement provided by the macroscopic YIG spin ensemble improves the projected sensitivity as
\(
N_s^{-1/2}
\)
at fixed $\chi$ and $\kappa_m$. Since $g_{mc}=g_0\sqrt{N_s}$ is itself a collective quantity, increasing $N_s$ at a fixed operating detuning $\Delta_{mc}$ also enhances the dispersive shift $\chi\propto g_{mc}^2$, so the realizable scaling is in fact stronger than $N_s^{-1/2}$, as shown explicitly below.

Because the transmon coherence times $T_1,T_2\sim100~\mu\mathrm{s}$, we thus also take a qubit coherence time $T_2=100~\mu\mathrm{s}$ and a measurement cycle time $t_{\rm cycle}=10\,T_2=1~\mathrm{ms}$, allowing $9\,T_2$ for state preparation, readout, and reset. Since the drive is extremely weak near the detection threshold, $1/(2\chi\langle\hat n\rangle)\gg T_2$, we fix $t_R=T_2=100~\mu\mathrm{s}$. We also take $N_s=1\times10^{21}$, with $\chi/2\pi=1.27~\mathrm{MHz}$ evaluated at the operating detuning $\Delta_{mc}/2\pi=5.7~\mathrm{GHz}$, corresponding to a Kittel-mode frequency $f_K=\omega_c/2\pi+\Delta_{mc}/2\pi\approx13.7~\mathrm{GHz}$, for which the Gilbert-damping linewidth [Eq.~\eqref{eq:GilbertDamping}] gives $\kappa_m/2\pi\approx2.6~\mathrm{MHz}$. Evaluating Eq.~\eqref{eq:OmegaMin} at a representative $T_{\rm int}=10^4~\mathrm{s}$, we obtain
\begin{equation*}
\Omega_{J,\min}^{(0)} \simeq 2\pi\times{1.16~{\rm kHz}}.
\end{equation*}
The corresponding projected sensitivity to the Majoron-electron coupling becomes
\begin{equation}
\left( \dfrac{c_e^A}{f_J}
	\right)_{\rm min}
	\simeq
2.7\times10^{-8}
	~{\rm GeV}^{-1}.
\end{equation}

The projected sensitivity reaches the
\(2.7\times10^{-8}\,\mathrm{GeV}^{-1}\)
level, highlighting the promise of hybrid magnon--qubit platforms as a new quantum-sensing architecture for ultralight Majoron dark matter.
By combining the collective enhancement of a macroscopic magnon mode with the phase sensitivity of superconducting-qubit
Ramsey interferometry, the proposed detector offers a complementary strategy for probing weak spin-dependent dark-matter interactions.

Figure~\ref{fig:snr}(a) shows the SNR as a function of the integration time $T_{\rm int}$ for a reference coupling strength $c_e^A/f_J=4.3\times10^{-9}~\mathrm{GeV}^{-1}$. As predicted by Eq.~\eqref{eq:SNR}, the SNR scales as $\sqrt{T_{\rm int}}$, appearing as a straight line with slope $1/2$ on the log-log scale. Comparing the resonant case ($\delta\tilde\omega_K=0$) with an off-resonant detuning $\delta\tilde\omega_K=\kappa_m/2$, the integration time $T_{\rm int}$ required to reach ${\rm SNR}=1$ increases from $1.6\times10^7~{\rm s}$ to $6.5\times10^7~{\rm s}$. Thus, the off-resonant case requires about four times longer integration than the resonant case. This behavior highlights the importance of operating near the resonance condition $m_J\simeq\tilde\omega_K$, where the signal is maximized and the measurement time is substantially reduced.

Figure~\ref{fig:snr}(b) shows the minimum detectable coupling $(c_e^A/f_J)_{\rm min}$ as a function of $T_{\rm int}$, for $N_s=1\times10^{21}$ and $N_s=1\times10^{22}$, with $g_{mc}$ (and hence $\chi$) rescaled for the larger spin number at the same operating detuning $\Delta_{mc}$. For both spin numbers, the sensitivity scales as $(c_e^A/f_J)_{\rm min}\propto T_{\rm int}^{-1/4}$, indicating that longer $T_{\rm int}$ provide only gradual improvements. Because $\chi\propto g_{mc}^2\propto N_s$ at fixed $\Delta_{mc}$, the two enhancements combine to shift the entire curve downward by a full order of magnitude, $(c_e^A/f_J)_{\rm min}\propto N_s^{-1}$, rather than the $N_s^{-1/2}$ expected from the collective drive alone. This gain comes at the cost of a narrower dispersive margin: $g_{mc}/\Delta_{mc}$ grows from $0.0995$ at $N_s=1\times10^{21}$ to $0.31$ at $N_s=1\times10^{22}$, so the leading-order dispersive approximation is less comfortably satisfied for the larger ensemble. Greater sensitivity gains can instead be achieved by enhancing the dispersive shift $\chi$ or extending $t_R$ toward the coherence limit. The weak dependence on $T_{\rm int}$ also motivates the mass-scan strategy discussed below.

\subsection{Mass scan strategy}

The projected sensitivity estimates presented in Sec.~\ref{sec:sensitivity}.B assume that the Majoron mass satisfies the resonance condition $m_J\simeq\tilde{\omega}_K$. Since the Kittel frequency is controlled by the static magnetic field, $\omega_K=\gamma_e B_0$, the resonance can be tuned continuously by varying $B_0$. This allows the same hybrid magnon--qubit detector to search for Majorons over a broad mass range.

For a scan covering frequencies between $f_{\rm min}$ and $f_{\rm max}$, the search range is divided into frequency bins of width $\kappa_m/2\pi$, determined by the Kittel-mode linewidth. The number of independent bins is

\begin{equation}
N_{\rm bins}
=
\int_{f_{\rm min}}^{f_{\rm max}}
\frac{df}{\kappa_m(f)/2\pi},
\label{eq:Nbins}
\end{equation}
and the integration time per bin is $T_{\rm int}=T_{\rm meas}/N_{\rm bins}$, where $T_{\rm meas}$ is the total measurement time. To model a realistic scan, we use the frequency-dependent magnon linewidth $\kappa_m(f)=2\pi f\alpha_G$ of Eq.~\eqref{eq:GilbertDamping}, with $\alpha_G=1.9\times10^{-4}$. Since the linewidth increases with frequency, the projected sensitivity gradually becomes weaker at larger Majoron masses.

\begin{figure}[t]
\centering
\includegraphics[width=\linewidth]{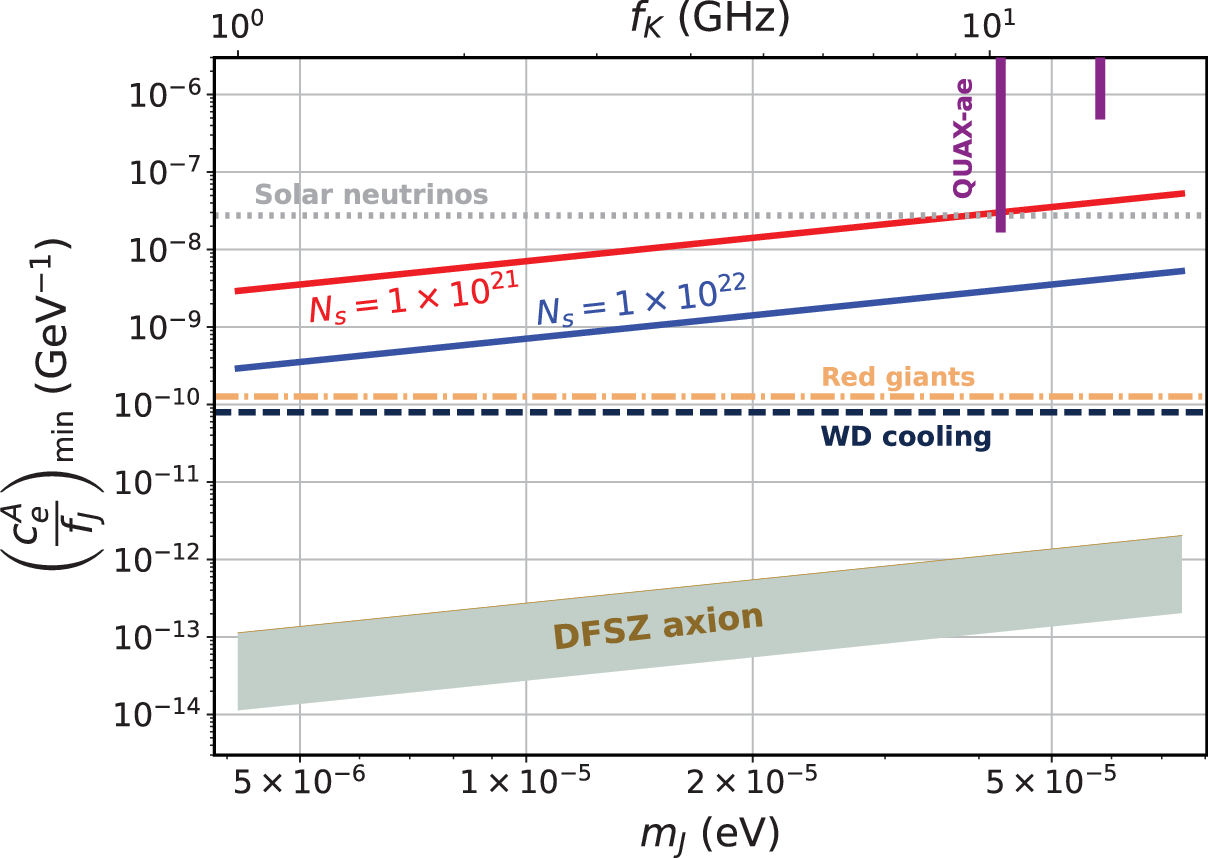}
\caption{Projected mass-scan sensitivity $(c_e^A/f_J)_{\rm min}$ versus Majoron mass $m_J$ (bottom axis) and Kittel/Majoron frequency $f_K=m_J/h$ (top axis), for  $T_{\rm meas} = 1$-yr, $\kappa_m(f)/2\pi=\alpha_G f$, $\alpha_G=1.9\times10^{-4}$, $t_R=100~\mu\mathrm{s}$. Filled purple bars mark the excluded segments of the QUAX ferromagnetic haloscope at $m_a\approx42.7~\mu\mathrm{eV}$~\cite{Crescini2020} and $m_a\approx58~\mu\mathrm{eV}$~\cite{Crescini2018}; the shaded diagonal band is the canonical DFSZ axion-electron prediction, $g_{aee}=2.8\times10^{-11}\cos^2\beta\,(m_J/{\rm eV})$ for $\cos^2\beta\in[0.1,1]$; dashed, dash-dotted, and dotted lines are the astrophysical/solar bounds from white-dwarf cooling in 47~Tucanae~\cite{Fleury2025}, the red-giant-branch tip~\cite{CapozziRaffelt2020}, and solar neutrinos~\cite{GondoloRaffelt2009}, respectively. All constraints are converted to $(c_e^A/f_J)_{\rm min}$ via $c_e^A/f_J\equiv g_{aee}/(2m_e)$, with couplings above each excluded.}
\label{fig:massscan}
\end{figure}

Figure~\ref{fig:massscan} shows the projected sensitivity for a one-year scan ($T_{\rm meas}$) over $f=1$--$18~\mathrm{GHz}$, corresponding to a Majoron mass range of $m_J\simeq4\times10^{-6}$--$7\times10^{-5}~\mathrm{eV}$. For a YIG sphere with $N_s=1\times10^{21}$, it reaches $(c_e^A/f_J)_{\rm min}\simeq2.9\times10^{-9}$--$5.3\times10^{-8}~\mathrm{GeV}^{-1}$. Increasing the spin number to $N_s=1\times10^{22}$ improves the sensitivity by $N_s^{-1}$, extending the reach to $(c_e^A/f_J)_{\rm min}\simeq2.9\times10^{-10}$--$5.3\times10^{-9}~{\rm GeV}^{-1}$. These results demonstrate that a year-long resonant scan with a larger YIG crystal can substantially enhance the discovery potential of the proposed hybrid magnon--qubit haloscope.

It is instructive to compare the projected sensitivity with existing constraints on the axion/Majoron--electron coupling. We use the relation $c_e^A/f_J\equiv g_{aee}/(2m_e)$ which connects our normalization to the standard convention adopted in the axion literature. Here, $g_{aee}$ denotes the dimensionless axion--electron pseudoscalar coupling, defined through the interaction $\mathcal{L}\supset (g_{aee}/2m_e)\,\partial_\mu a\,\bar e\gamma^\mu\gamma^5 e$, where $m_e$ is the electron mass.

A useful benchmark is the QUAX-ae ferromagnetic haloscope, which has directly searched for this coupling using a YIG sphere ensemble with $N_s=1.0\times10^{21}$ spins~\cite{Crescini2018,Crescini2020}. The experiment reported limits of $(c_e^A/f_J)_{\rm min}\approx1.7\times10^{-8}$ at $m_a=42.7~\mu\mathrm{eV}$ and $4.8\times10^{-7}~\mathrm{GeV}^{-1}$ at $m_a=58~\mu\mathrm{eV}$, shown as filled bars in Fig.~\ref{fig:massscan}. For a direct comparison, we adopt the same total spin number and model it as a single YIG sphere to maintain the single-Kittel-mode description. Under these conditions, our benchmark sensitivity of $(c_e^A/f_J)_{\rm min}\approx2.7\times10^{-8}~\mathrm{GeV}^{-1}$ at $T_{\rm int}=10^4~\mathrm{s}$ is within a factor of $2$ of the strongest QUAX-ae limit and about $18\times$ stronger than the limit at $58~\mu\mathrm{eV}$, despite using an integration time roughly $50\times$ shorter. This comparison demonstrates that QND qubit readout is broadly competitive with conventional linear-amplifier detection at the same spin-ensemble size.

The closest architectural precedent is the magnon-QND experiment of Ikeda~\textit{et al.}~\cite{Ikeda2022}, which reported $g_{aee}<2.6\times10^{-6}$, corresponding to $(c_e^A/f_J)_{\rm min}\approx2.5\times10^{-3}~\mathrm{GeV}^{-1}$. Compared with this result, our benchmark sensitivity is improved by roughly five orders of magnitude, primarily due to the larger spin ensemble and the collective Ramsey-based readout strategy.

Finally, we compare with model-independent stellar and solar energy-loss bounds. Solar neutrino measurements constrain the coupling to $(c_e^A/f_J)\lesssim2.8\times10^{-8}~\mathrm{GeV}^{-1}$~\cite{GondoloRaffelt2009}, while more stringent limits of $(c_e^A/f_J)\lesssim1.3\times10^{-10}$ and $8\times10^{-11}~\mathrm{GeV}^{-1}$ are obtained from red-giant observations and white-dwarf cooling in the globular cluster 47~Tucanae, respectively~\cite{CapozziRaffelt2020,Fleury2025}. Reaching these sensitivities would require further increases in $N_s$, $\chi$, or the coherence time $T_2$ beyond the values considered here.
Direct laboratory searches therefore provide a complementary probe that is independent of astrophysical modeling uncertainties and is performed under well-controlled terrestrial conditions.

\section{Conclusions}
\label{sec:conclusion}
We have proposed a hybrid magnon--qubit approach for detecting ultralight Majoron dark matter through its derivative coupling to electron spins. The oscillating Majoron background acts as an effective pseudo-magnetic field that resonantly drives the Kittel mode of a YIG sphere, producing a collective magnon response enhanced by the large spin ensemble. By coupling the magnon mode dispersively to a superconducting transmon qubit through a microwave cavity, the resulting magnon population can be converted into a qubit-frequency shift and measured using QND Ramsey interferometry.

Our analysis shows that this hybrid architecture combines collective spin enhancement with high-fidelity quantum readout, enabling sensitive searches for spin-coupled Majoron dark matter. The magnetic-field tunability of the Kittel mode further allows the resonance condition to be scanned, giving access to a broad Majoron mass range across a family of cavity-qubit stages tuned to different frequency bands.
These results establish hybrid magnon--qubit platforms as a promising interface between quantum sensing and dark-matter detection. Looking forward, extending the present single-qubit scheme to networked quantum architectures may provide a route toward exploiting collective quantum resources for enhanced searches for Majoron dark matter and other ultralight bosonic fields.

\section*{Acknowledgments}
This work was supported by the Vietnam Academy of Science and Technology (VAST) under Grant No.~CBCLCA.03/25-27, the Tohoku Initiative for Fostering Global Researchers for Interdisciplinary Sciences (TI-FRIS) under MEXT's Strategic Professional Development Program for Young Researchers, and FRIS Creative
Interdisciplinary Collaboration Program.

\section*{Data availability}
The data that support the findings of this article are
openly available at~\cite{code4majoron}.

\end{document}